# Breaking the Mode Degeneracy of Surface-Plasmon Resonances in a Triangular System

Nahid Talebi*[a], Wilfried Sigle[a], Ralf Vogelgesang[d], Christoph T. Koch[a,b], Cristina Fernández-López,
Luis M. Liz-Marzán[c], Burcu Ögüt[a], Melanie Rohm[a], Peter A. van Aken[a]

[a]Max Planck Institute for Intelligent Systems, Heisenbergstraße 3, D-70569 Stuttgart, Germany

[b]Ulm University, Albert-Einstein-Allee 11, D-89081 Ulm, Germany

[c]Departamento de Química Física, Universidade de Vigo, 36310, Vigo, Spain

[d]Max Planck Institute for Solid State Research, Heisenbergstraße 1, D-70569 Stuttgart, Germany

Corresponding author:

*E-mail: talebi@is.mpg.de; Tel. +49 711 689 3649; Fax +49 711 689 3522.

**ABSTRACT:** In this paper, we present a systematic investigation of symmetry-breaking in the plasmonic modes of triangular gold nanoprisms. Their geometrical $C_3$ symmetry is one of the simplest possible that allows degeneracy in the particle's mode spectrum. It is reduced to the non-degenerate symmetries $C_v$ or E by positioning additional, smaller gold nanoprisms in close proximity, either in a lateral or a vertical configuration. Corresponding to the lower symmetry of the system, its eigenmodes also feature lower symmetries ($C_v$), or preserve only the identity (E) as symmetry. We discuss how breaking the symmetry of the plasmonic system not only breaks the degeneracy of some lower order modes, but also how it alters the damping and eigenenergies of the observed Fano-type resonances.

## INTRODUCTION

Metallic triangular nanoparticles can support various localized surface plasmon resonances (LSPR). The basic resonance modes have field maxima at the corners, at the center of the side faces, and in the center of the triangle.[1] Using the discrete-dipole approximation, Shuford et al.[2] found that increasing the edge length and/or decreasing the prism height produces more intense red shifted dipole peaks which was later on experimentally confirmed by Nelayah et al.[3] Shuford et al. also identified dipolar and quadrupolar modes. Experimental evidence for multipolar modes was given by Gu et al.[4] who found modes up to the 5$^{th}$ order. With the use of finite-difference time-domain (FDTD) calculations they could show that it is wedge plasmons that give rise to the multipolar resonances with field maxima along the triangle edges. Furthermore, it was shown that the resonances form by interference of plasmons propagating around the triangle along the wedges and plasmons reflected at the triangle corners.

Symmetry breaking in a spherical system was studied by Wang et al.[5] showing the modification of selection rules. There are indications of symmetry breaking in triangular systems in References 1 and 6. In chemistry, the relation of structural and orbital field symmetry is often used to predict the evolution of spectral properties. This approach has recently been transferred to the study of strongly coupled nano-particles and identical hybridized nano-systems, such as plasmonic oligomers[7,8,9]. An excellent survey of this field is reported in Ref. 10.

Here we systematically reduced the symmetry of a triangular system by studying LSPRs of coupled dissimilar triangles, i.e. of triangles that are close to each other laterally or on top of each other. The presence of a second triangle reduces the symmetry of the system which in turn influences the plasmonic response. Plasmon resonances that are degenerate in a highly symmetric particle split into non-degenerate modes by breaking the symmetry. Understanding symmetry breaking effects may become an effective way of designing plasmonic systems for manipulation of the transmission of light.

## EXPERIMENTAL AND CALCULATION DETAILS

**Synthesis of Gold Nanoprisms.** Tetrachloroauric acid (HAuCl$_4$×3H$_2$O) and sodium borohydride (NaBH$_4$) were purchased from Aldrich. Ascorbic acid, citric acid (C$_6$H$_5$O$_7$Na$_3$× 2H$_2$O), potassium iodide (99%) and sodium hydroxide (97%) were supplied by Sigma. Cetyltrimethylammonium bromide (CTAB) was procured from Fluka. All chemicals were used as received. Milli-Q grade water was used in all preparations. The synthesis of Au nanoprisms was based on a seeded growth method.[11] Au nanoparticle seeds (average diameter ≈ 5 nm) were prepared by adding NaBH$_4$ (100 mM, 1 mL) to HAuCl$_4$ aqueous solution (0.27 mM, 37 mL) containing 10$^{-5}$ moles of sodium citrate. After 5 min under vigorous stirring, the mixture was allowed to react overnight to allow the complete hydrolysis of unreacted NaBH$_4$. Three growth solutions were prepared for the seed-mediated growth steps. The first two solutions (A and B) were identical and contained HAuCl$_4$ (10 mM, 0.25 mL), NaOH (0.1 M, 0.05 mL), ascorbic acid (0.1 M, 0.05 mL) and CTAB (0.05 M, 9 mL; also containing 75 µL of 0.1 M KI). The final growth solution (C) comprised $HAuCl_4$ (10 mM, 2.5 mL), NaOH (0.1 M, 0.50 mL), ascorbic acid (0.1 M, 0.50 mL), and CTAB (0.05 M, 90 mL; with 75 µL of 0.1 M KI). Particle formation was initiated by adding 1 mL of seed solution to growth solution A. The solution was gently shaken and then 1 mL of this mixture was immediately added to growth solution B. After shaking, all of the resulting mixture was added to solution C. After the final addition, the color of C changed from clear to purple over a period of 30 minutes.

**Transmission electron microscopy.** For imaging and electron energy-loss spectroscopy (EELS) a Zeiss SESAM microscope[12] was used. This instrument is equipped with an omega-type electron monochromator which allows an energy resolution below 100 meV. After passing the thin sample, electrons exhibit a quasi-continuous energy-loss spectrum, dominated by bulk plasmon excitations, but also excitations due to surface plasmons, interband transitions, core-loss excitations, etc. Electrons of different energy are dispersed in the in-column MANDOLINE energy filter[13] resulting in an energy-loss spectrum in its energy-dispersive plane. Using a mechanical slit (0.2 eV energy width in our case) electrons of a particular energy loss can be selected and an image is formed only from these electrons. This is known as energy-filtering TEM (EFTEM). By changing the high voltage of the microscope the selected energy loss is varied, in our case from 0 to 4 eV. This ensures that the electrons passing the energy filter have constant energy and no degradation of the optical properties of the filter takes place. In fact, EFTEM image series are always recorded from high to low energy loss because image intensity increases dramatically towards 0 eV energy loss. By doing so, image artifacts from scintillator afterglow are avoided. The resulting EFTEM image series can be considered as a 3D data cube with two spatial dimensions and an energy axis. The EFTEM images are recorded on a 2048×2048 pixel CCD camera. For EFTEM, the camera is typically binned by a factor of 4, resulting in 256×256-pixel images, in order to reduce noise. Spatial resolution is not affected by binning because of the limitations of resolution mentioned below. Complementary to EFTEM, EELS spectra can be recorded from small areas by focusing the electron beam on a sub-nm spot. Typical spectra of 2048 channels cover an energy range of about 20 eV. Line scans or 2D maps (EELS spectrum images, EELS-SI) are obtained by scanning the spot across the desired line (area). In order to have a reliable energy calibration, energy-loss spectra are recorded including the elastic peak at 0 eV. Because this is orders of magnitude stronger than the plasmon peaks of interest, the EELS spectrum is spread along the non-dispersive direction of the CCD camera. This prevents saturation of the camera pixels. EFTEM and EELS-SI are complementary in the sense that EFTEM offers high spatial sampling and thus gives a fast overview of the spatial distribution of resonances, whereas EELS-SI gives better sampling and better signal-to-noise ratio in energy space. EELS-SI is therefore better suited to separate fine spectral details.

**Spatial resolution.** Electron beams can nowadays be focused to sub-Å spots using aberration-corrected electron optics. This allows obtaining high-angle annular dark-field images of sub-Å resolution. Close to 1 Å resolution is also achieved using electrons that have excited deeply bound 1s- or 2p-electrons thus providing atomically resolved elemental maps. However, at small energy losses the spatial resolution is degraded due to the long-range Coulomb force between the electron probe and the dielectric material. This can be understood from image formation theory.[14] A 200-keV-electron having suffered 5 eV energy loss has a characteristic inelastic scattering angle of only 12.5 µrad. According to the Rayleigh criterion small scattering angles are related with poor spatial resolution. In the energy-loss range of interest for surface plasmons (0–5 eV) the spatial resolution is of the order of 5–10 nm.[4,10] This obviously far exceeds the electron probe size and is in the range of the latest scanning near-field optical microscopy techniques. An important advantage of electron spectroscopy is that energy losses from about 400 meV ($\lambda = 3$ µm) up to several keV ($\lambda = 5$ Å) can be detected. This of course covers the whole optical spectrum. Despite electron monochromation, below 400 meV spectral details start to become invisible because of the overlap with the tail of the elastic peak at 0 eV.

**Peak-finding algorithm.** EFTEM is a highly efficient technique for the fast visualization of the spatial distribution of particular energy losses. However, in the regime of low energy losses it reaches its limits when weak energy-loss features superimposed by a large background are to be displayed. This is because background subtraction is complicated due to the lack of an analytical background function, as is available at high energy losses. Therefore we developed a simple peak-finding algorithm which allows displaying the spatial locations of weak spectral peaks. The algorithm was written in Digital Micrograph (Gatan, USA) script language. For this, spectra are extracted from each image pixel ($i, j$) ($i, j = 1…256$) of the EFTEM series data cube. In each spectrum, those energy positions are set to intensity '1' whose neighbouring energy channels have lower intensity. All other energy values are set to intensity '0'. This results in a new data cube with binary data. Cutting this new cube parallel to the $x$–$y$-plane directly displays the spatial distribution of spectral peaks. Peaks below 1 eV energy loss would normally not be detectable by this technique because they overlap severely with the tail of the elastic peak at 0 eV and therefore appear as a shoulder rather than as a peak. In order to identify also these peaks we eliminated the zero-loss tail by extrapolation and subtraction of a power-law function fitted to a region of the elastic peak between 0.3 and 0.45 eV.

**FDTD calculations.** Using the finite-difference time-domain method, the electromagnetic response of the structure was theoretically investigated. In order to effectively resolve different modes of the structure, a symmetry decomposition technique was exploited. Considering a simple triangle with three symmetry planes, each mode can be distinguished by being either symmetric or anti-symmetric with respect to a specific symmetry plane. Anti-symmetric modes are distinguished by considering a perfect electric conductor (PEC) at the position of a symmetry plane, while the symmetric modes hold true by a perfect magnetic conductor (PMC) positioned at the same symmetry plane. In our analysis, we have used different polarizations of the excitation beams to satisfy the pre-assumed configurations of either PEC or PMC symmetry planes.

For our FDTD analysis, the computation domain was discretized by cubic cells of 5 nm side length and the normal Yee's mesh was exploited. To model the dispersive behavior of gold, the Drude model with two added critical-point functions was used.[15] A Ricker-wavelet function was introduced as the temporal dependence of the incident beam, which allows us to excite energetically different modes of the structure in only one simulation attempt. By applying a discrete Fourier transform technique, the near-field spectrum of the structures was obtained. We used the concept of an additive source in our code, which allows us to introduce a vibrating plane at a location of about 1 μm above the structure. The spatial distribution and the temporal delays at the different points of the excitation plane are controlled in a way to preserve the desired optical beam, so that either linearly polarized plane-waves or TE and TM modes of the free space are excited. The TE mode is the free-space mode of electromagnetic waves having no $z$-component of the electric field, while the TM mode has no $z$-component of the magnetic field.

## RESULTS AND DISCUSSION

Figure 1 shows the bright-field TEM images of the structures that are investigated in this paper. The single triangle shown in Fig. 1a exhibits 3 symmetry planes. The structure shown in Fig. 1b is composed of a small triangle of only 120 nm edge length on top of a larger triangle of about the same size as that of the single triangle. The location of the small triangle is such that the structure has still one symmetry plane. The structure shown in Fig. 1c has no symmetry plane. The height of the large and small triangles, determined from the low-loss EELS data, is 40 nm and 30 nm, respectively.

Using the FDTD method, the optical modes of the simple Au nanoprism were numerically investigated. In order to determine the resonance energies, the distribution of the near-field intensity of the electric field along the edges of the nanoprism was computed. Figure 2 shows the various computed modes for the different configurations of symmetry planes. The middle column shows the intensity of the electric field computed at the location of the test points, positioned along the vertical edge of the triangle. The fundamental mode of the single triangle is the doubly degenerate E-mode of the $C_{3v}$ group.[16,17] There is no difference in energy between the PMC and PEC symmetry plane configurations, although a distinct behavior is obvious for the spatial distribution of the field components. The right column shows the spatial distribution of the $z$-component of the electric field at the resonance energies, at 5 nm above the nanoprism. These very different modes can be excited optically by the two orthogonal linear polarizations, respectively. Nevertheless, their resonance energies are identical. An intuitive understanding emerges if one considers the alternative circular symmetry. This kind of triangle does not distinguish between clockwise and anti-clockwise direction. Correspondingly, optical absorption spectra using left- and right-circularly polarized light must reveal identical peak energies. As the linear polarization states can be constructed by superposition of the two circular ones, all their resonance energies are the same.

For an incident TM wave, the structure sustains two plasmonic resonances, distinguished by two peaks in the intensity spectrum. Both belong to the $A_1$-symmetry class of the $C_{3v}$ group. The lower-energy mode with the hot spots mainly at the center of the edges is a monopolar mode, while the higher-energy one is a multipolar mode with a hot spot at the center of the triangle. These modes are denoted by III and IV, respectively. Mode III has a broad spectrum and correspondingly a short dephasing, while mode IV has a narrow spectrum. This situation is appropriate for the excitation of Fano-type resonances[18]. However, in order to see this kind of resonance, an observable is required that is sensitive to the interference[18]. It is observed in the near-field spectrum at the location of some precisely defined test-points. Figure 3 shows the calculated near-field spectrum at the location of the test point depicted at the inset. The sharp dip at 1.6 eV is due to the interference of the modes III and IV.

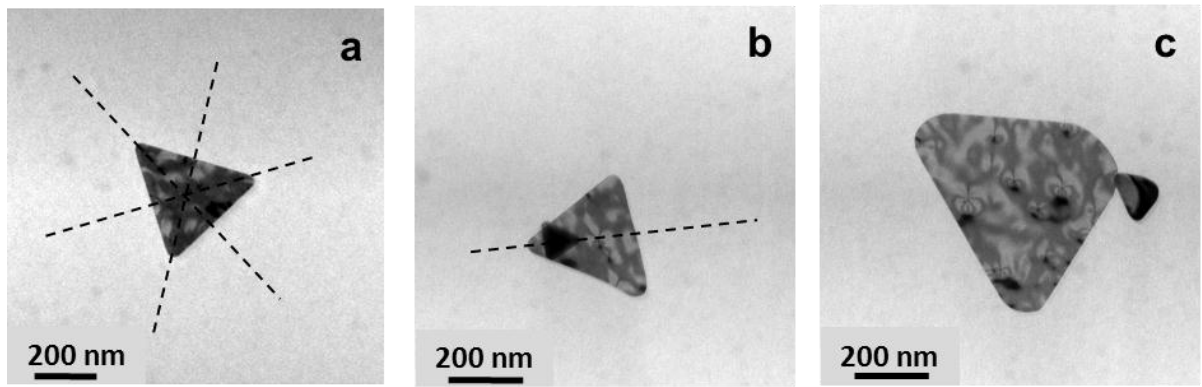

**Figure 1.** Bright-field images of (a) a single Au triangular nanoprism (edge length 400 nm), (b) a small Au triangle of 120 nm edge length on top of a larger triangular nanoprism with 370 nm edge length, and (c) a small Au triangle of 120 nm edge length located close to the corner of a larger nanoprism with 500 nm edge length. All nanoprisms are deposited on a 30 nm thick $SiN_x$ membrane.

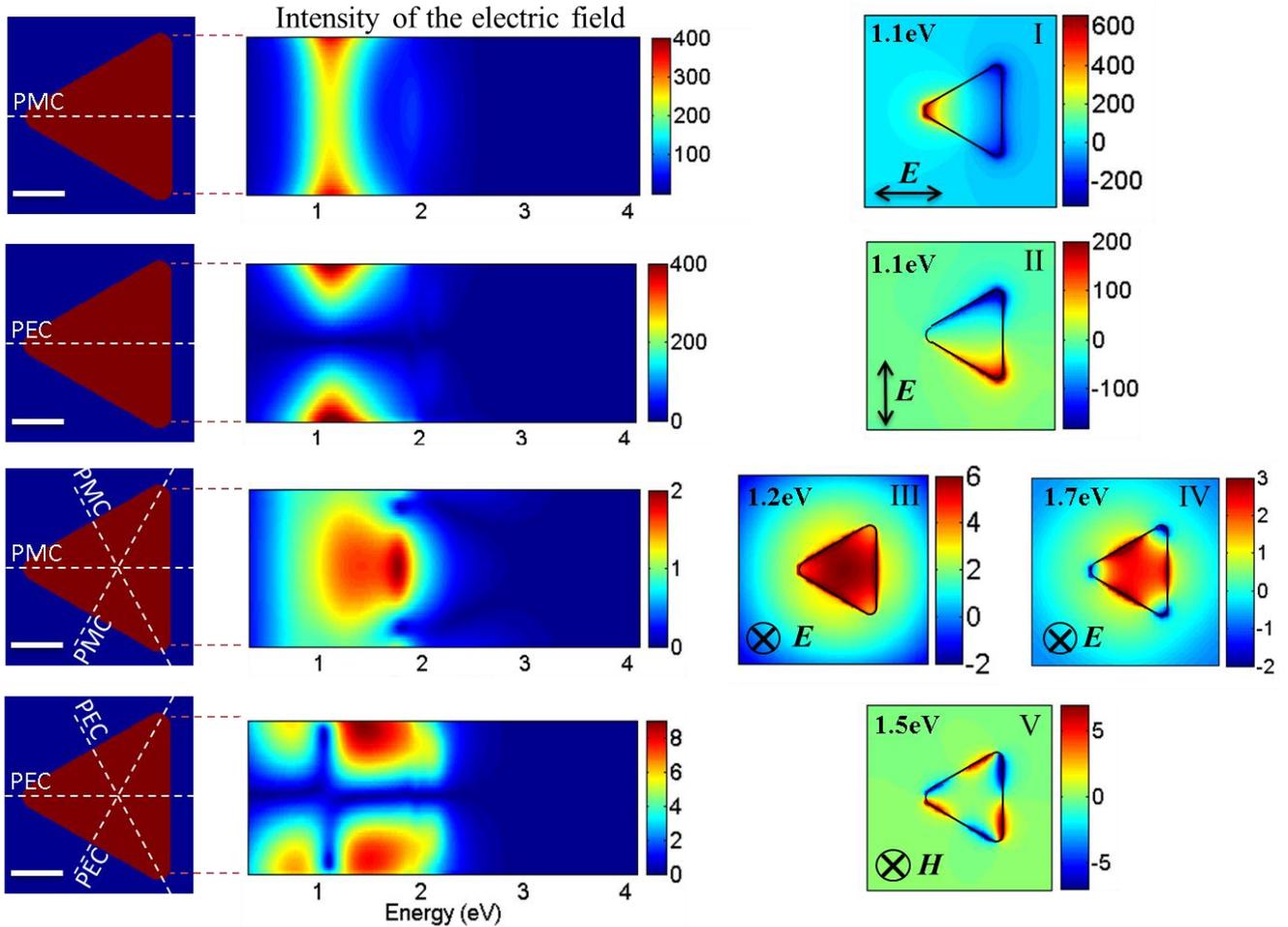

**Figure 2.** Plasmonic resonances of the single triangle, as calculated by FDTD, distinguished by their different symmetries. The middle column shows the intensity spectrum along the right edge. The right column shows the spatial distribution of the *z*-component of the electric field in the *xy*-plane at a distance of 5 nm above the nanoprism.

The experimental EELS and EFTEM results of this structure are shown in Figure 4. From the EFTEM data (Figure 4a–c) modes I/II, III, and V (which has $A_2$-symmetry) are clearly distinguishable. However, detection of mode IV (Figure 4g) was only possible using the peak maps (Figure 4d–g) which show all monopolar and multipolar modes of the triangle. It should be noted that the modes shown as I and II in Figure 2 are degenerate and not distinguishable in the EFTEM images. It should also be noted that energy loss values from EFTEM data have an error of ± 0.1 eV because of the slit width of 0.2 eV.

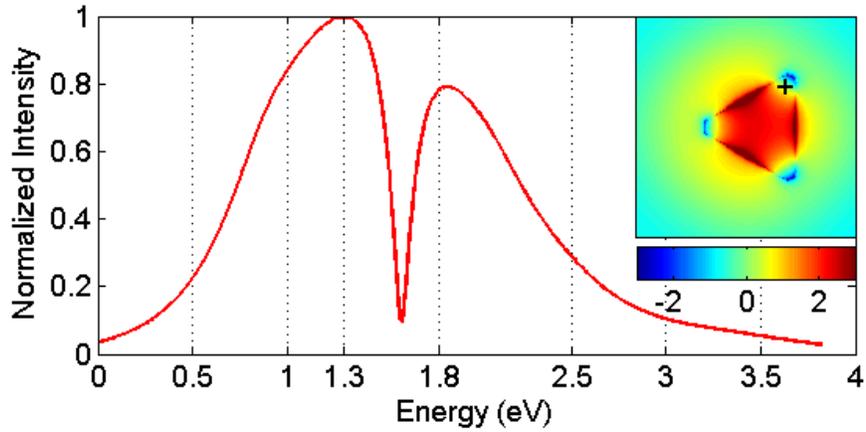

**Figure 3**. Fano resonance due to the overlap of the broad and narrow optical modes of the simple nanoprism, calculated at the location of the test point depicted at the inset. The inset shows the spatial distribution of the z-component of the electric field, at the energy of 1.8eV.

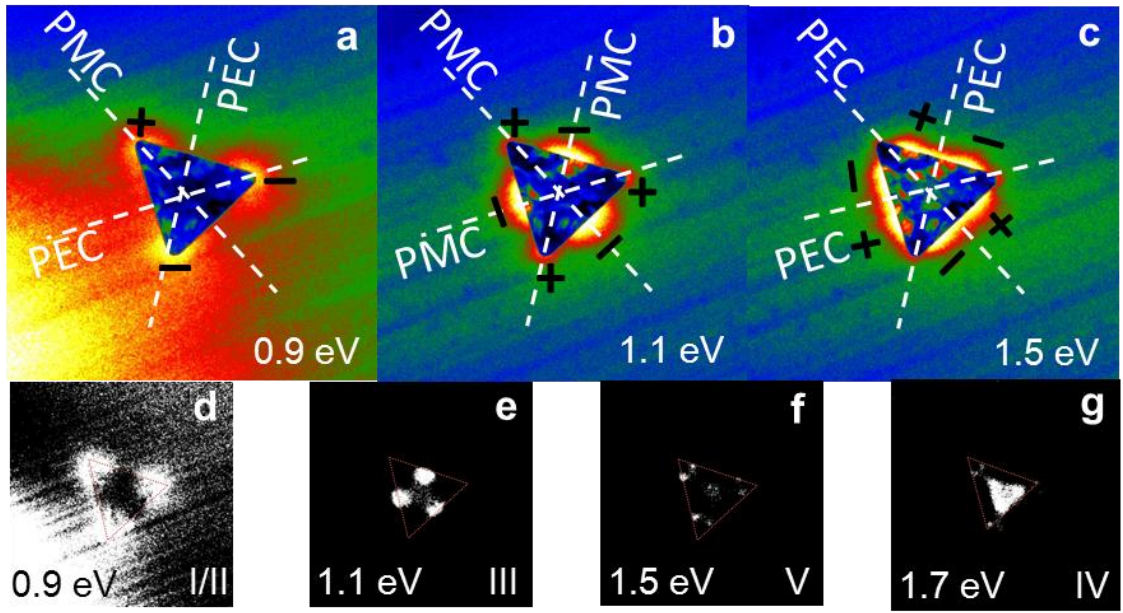

**Figure 4.** (a–c) EFTEM images and (d–g) peak maps of a single triangular nano-prism at the depicted energy losses. Roman numbers denote the modes as specified in Figure 2.

The Fano resonance observed in the simulation by decomposing the optical modes into the symmetric and anti-symmetric ones cannot be observed in the EELS analysis of the simple nanoprism. This is because the passing fast electron can excite any mode with non-vanishing $E_z$-nearfield component. An EFTEM image displays the incoherent superposition of electrons having excited any eigenmode of the structure with an eigenenergy within the range of the energy-selecting slit. Especially modes which are degenerate cannot be distinguished. Mode V has a resonance energy exactly located at the dip of the Fano resonance. The overlap of the spectra due to the excitation of all the modes will not allow an observer to report the interference occurring by the two resonances of III and IV.

The 3-fold $C_{3v}$ symmetry of the Au nanoprism is reduced to $C_v$ by positioning a smaller nanoprism symmetrically on top of the large nanoprism, as shown in Figure 1b. This structure has only one symmetry plane, which classifies all eigenmodes into either A' (symmetric) or A'' (anti-symmetric). Figure 5 shows the computed eigenmodes of this structure. It is obvious that modes I and II are still nearly degenerate, even though they suffered breaking of their former E-symmetry into A' (mode I) and A'' (mode II). Apparently, the symmetry breaking potential of the small add-on particle is not large enough to induce significant energy splitting.

Another A' eigenmode is observed at an energy of 1.9 eV. This originates from the coupling of the fundamental "corner mode" of the small triangle with an induced dipolar mode of the large triangle. It is important to note that this induced dipolar mode is not present in a single triangle at this eigenenergy.

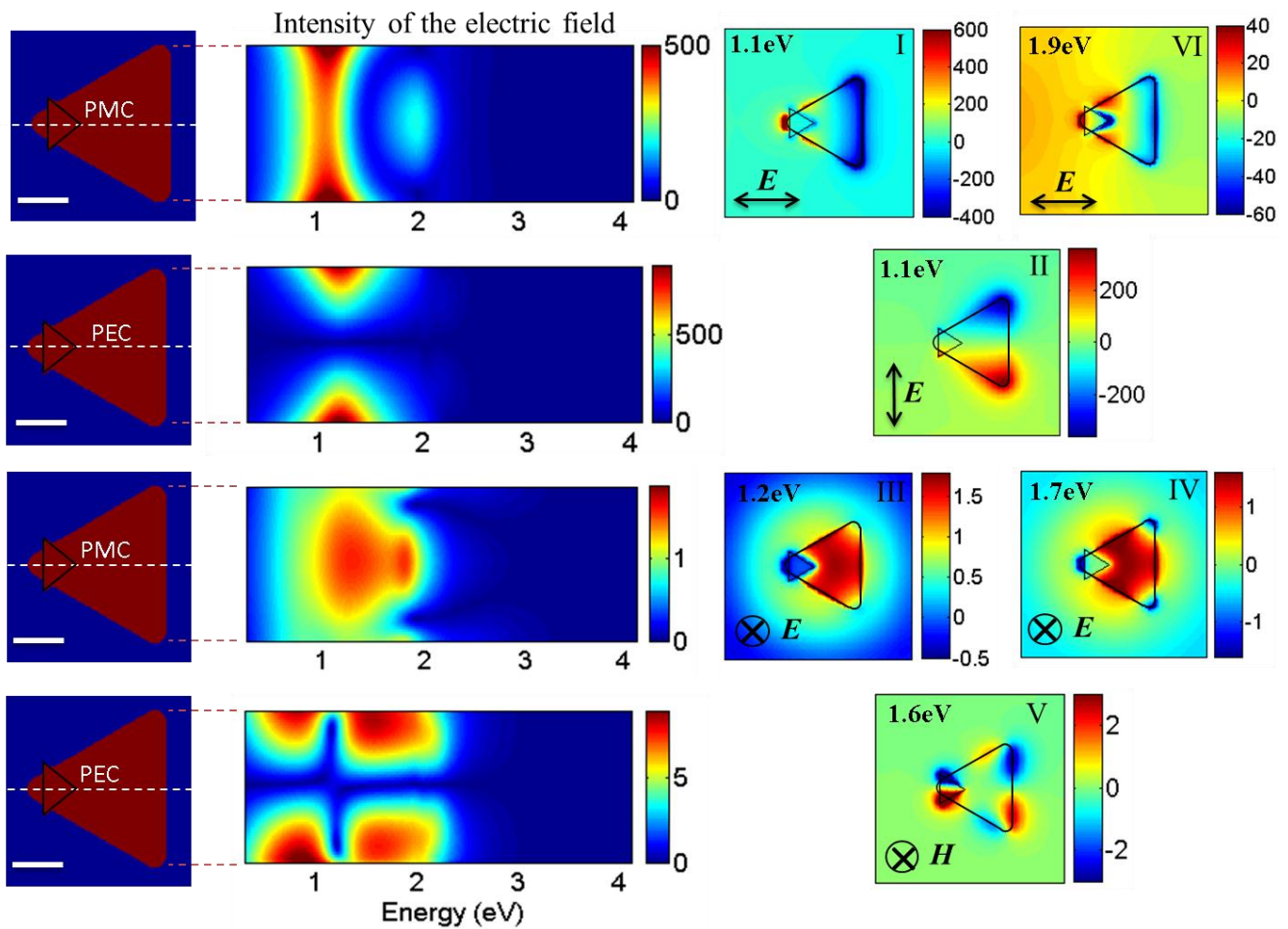

**Figure 5.** FDTD computed plasmonic resonances of the coupled triangles (Figure 1b) distinguished by their different symmetries. The middle column shows the intensity spectrum at the position of the test points shown in the left image. The right column shows the spatial distribution of the z-component of the electric field in the xy-plane at a distance of 5 nm above the nanoprism.

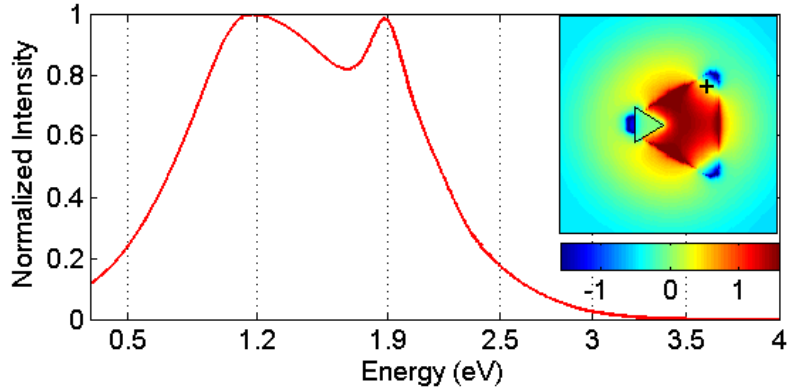

**Figure 6**. Fano resonance due to the overlap of the broad and narrow optical modes of the coupled nanoprisms, observed at the location of the test point depicted in the inset. The inset shows the calculated spatial distribution of the z-component of the electric field, at an energy of 1.7eV. This figure should be compared with Figure 3.

When comparing mode IV of this structure with the corresponding mode of the simple nanoprism, a broader resonance is observed for this structure. This is especially evident from the intensity spectrum. This phenomenon will result in a much narrower dip at the position of the test point depicted in the inset of Figure 3. The Fano-type resonance at the location of this test point is shown at Figure 6. It is evident that this dip is shallower in comparison with that of Figure 3. Moreover, the peaks are also shifted by 1eV.

Figure 7a shows a series of 50 EELS spectra acquired along the dashed arrow in the left image. Three resonance peaks, denoted by A, B, and C, can be distinguished at energy losses of 0.75 eV, 1.2 eV, and 1.8 eV. The corresponding EFTEM and peak maps at the identified eigenenergies (Figure 8) show a good correspondence to the FDTD data. Symmetry breaking is readily visible from Figure 8e. Careful inspection of Figure 7a shows that the energy of peak C red-shifts from the corner of the small triangle towards the edge of the large triangle. This can be attributed to the simultaneous excitation of modes V (1.7 eV) and VI (1.9 eV) by the fast electron. Moreover, the tail of the broad mode labeled as III can also interfere with the mentioned resonances. By fitting Gaussian functions to the 50 spectra this energy shift as well as the peak width were precisely determined (Figure 7b). Obviously the width of peak C exhibits a maximum near the sharp corner of the small triangle which we believe is due to radiation losses.

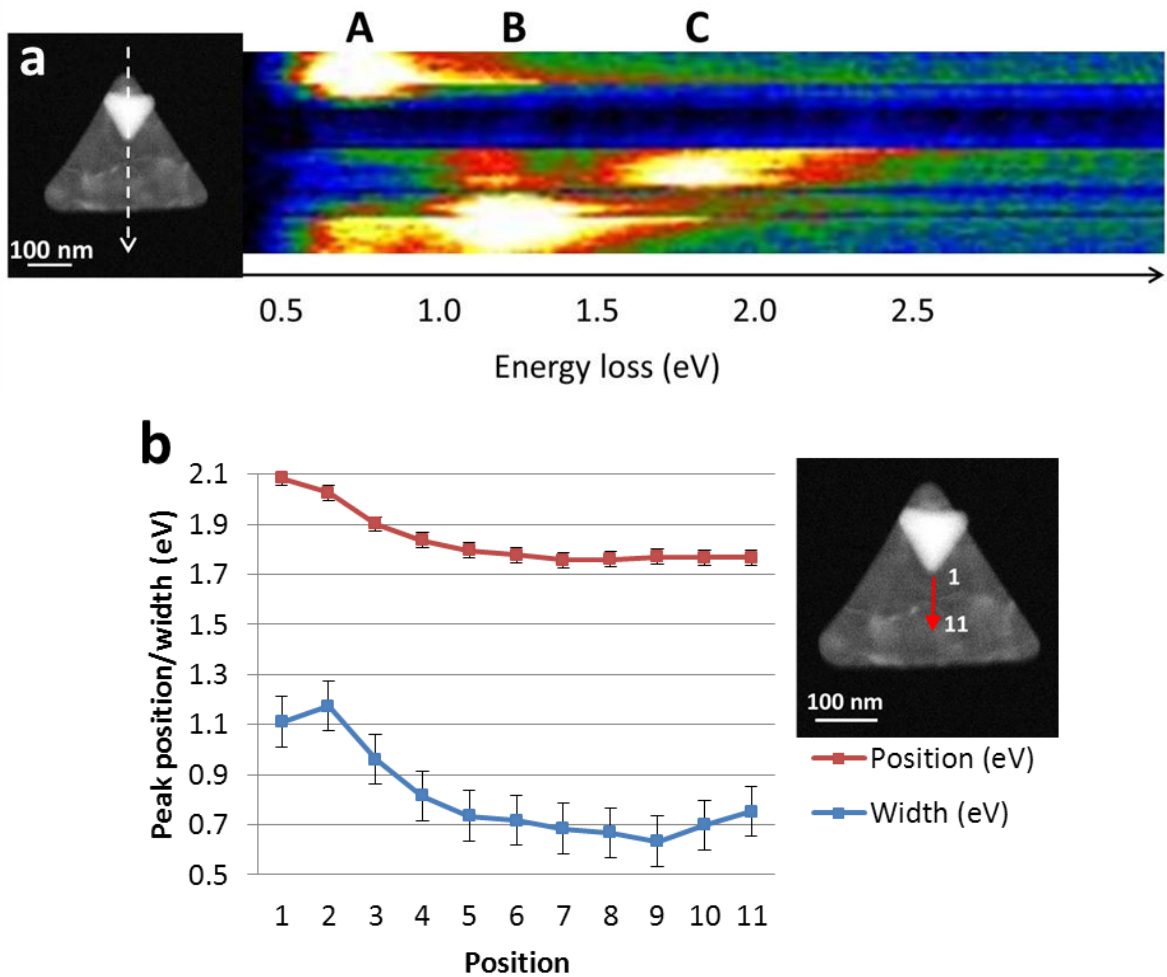

**Figure 7.** (a) 50 electron energy-loss spectra acquired along the arrow depicted in the left image. Three peaks at 0.75 eV (A), 1.2 eV (B), and 1.8 eV (C) are discernible. Peak A is excited close to the corner, peak B in the center, and peak C at the edge of the large triangle, compare also with Figure 8. (b) Dependence of position and width of peak C on the position of electron excitation, as indicated by the red arrow on the image at the right.

The damping of the plasmonic resonances can be readily computed from the width of the resonances. Moreover, the quality factor can be estimated as the ratio of the resonance frequency over the full width at half maximum of a Gaussian resonance. To achieve this, a summation of the Gaussian resonances is fitted to the spectrum, for both numerical and experimental results. Using this procedure, we have obtained the quality factor of the optical resonances shown in Figures 5 and 8. The highest quality factor corresponds to the optical mode IV which equals to 7.92, while mode III has the lowest quality factor of only 1.20.

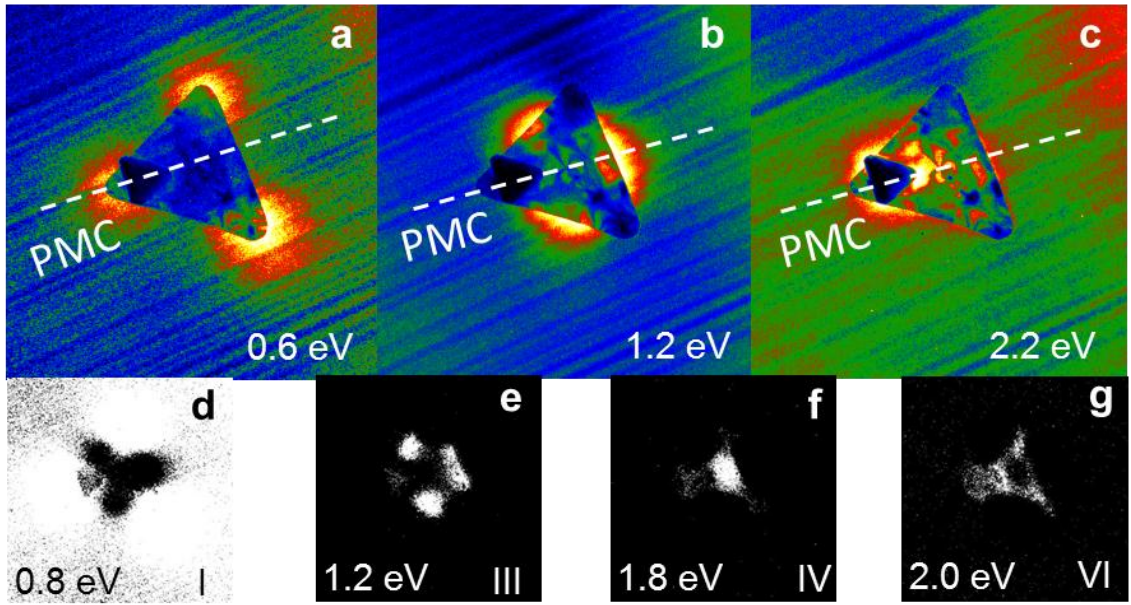

**Figure 8.** (a–c) EFTEM images of a small triangular Au nanoprism (edge length 120 nm) on top of a large Au nanoprism (edge length 370 nm). (d–h) peak maps at different energy losses, extracted from the EFTEM series. Roman numbers denote the modes as specified in Figure 5.

The position of the small triangle on top of the Au nanoprism is such that it does not disturb the mode profiles of the large nanoprism, so their coupling is considered to be weak. This is the reason for the good agreement between the resonance energies of the structures depicted in Figure 1a and Figure 1b. The position of the smaller triangle shown in Figure 1c is such that the overall structure preserves no symmetry, that is, the appropriate symmetry group is E, which contains only the identity. The results of numerical investigations for the plasmonic resonances of this structure are shown in Figure 9. It should be mentioned that the size of the larger nanoprism in the simulation is assumed to be the same as for the previous structures, for the sake of an easier comparison between their resonance energies, and to show the effect of the symmetry breaking on the plasmonic resonances in a distinct way. As it is obvious from these results, the modes numbered as I and III are not degenerate anymore. Moreover, in comparison with the previous structures, these modes have exchanged their sequence with regard to their energies and the incident polarization. Similar to the previous structure, the fundamental mode of the small triangle is obvious at 1.9 eV.

The Fano-type resonances which were observable at the previous structures are no longer preserved here, which is directly related to the symmetry breaking phenomenon which has a strong effect on the distribution and the quality factor of the mode labeled as III.

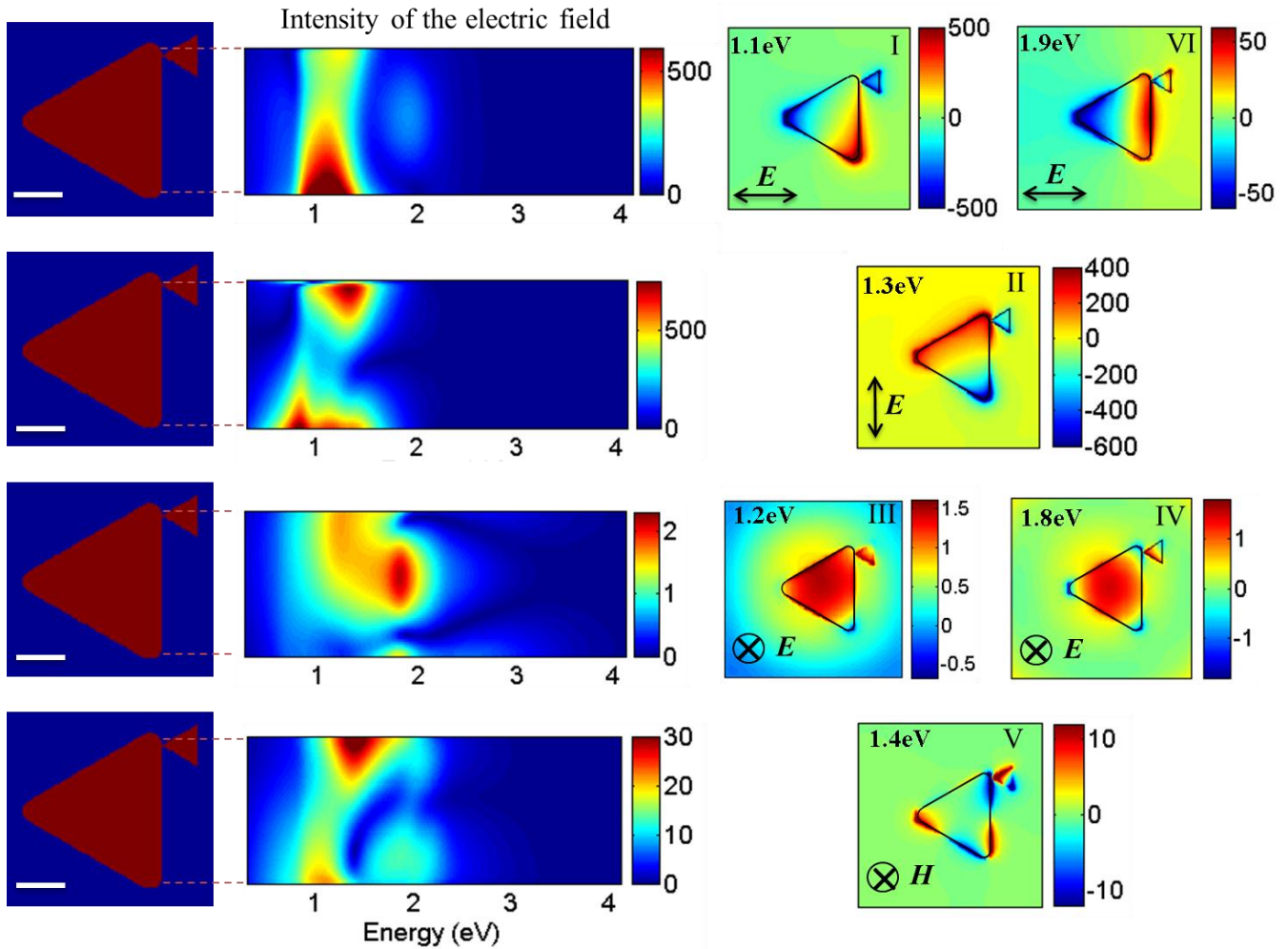

**Figure 9.** FDTD computed plasmonic resonances of the coupled triangles (Figure 1c). The middle column shows the intensity spectrum at the position of the test points depicted in the left image. The right column shows the spatial distribution of the *z*-component of the electric field in the *xy*-plane at a distance of 5 nm above the nanoprisms.

Figure 10 shows the EFTEM and peak maps of the structure shown in Figure 1c. It should be noted that the symmetry of the simple nanoprism is not only broken by the smaller nanotriangle, but also because of the rounder upper corner of the large nanoprism. This may result in some degree of deviation of the numerical results versus the experimental ones. The lowest-energy resonance (Figure 10a) is hardly visible because of the overlap with the tail of the strong elastic zero-loss peak. This energy is not present in the peak map plots (Figure 10 d-g). The other resonances are in qualitative agreement with our numerical simulations.

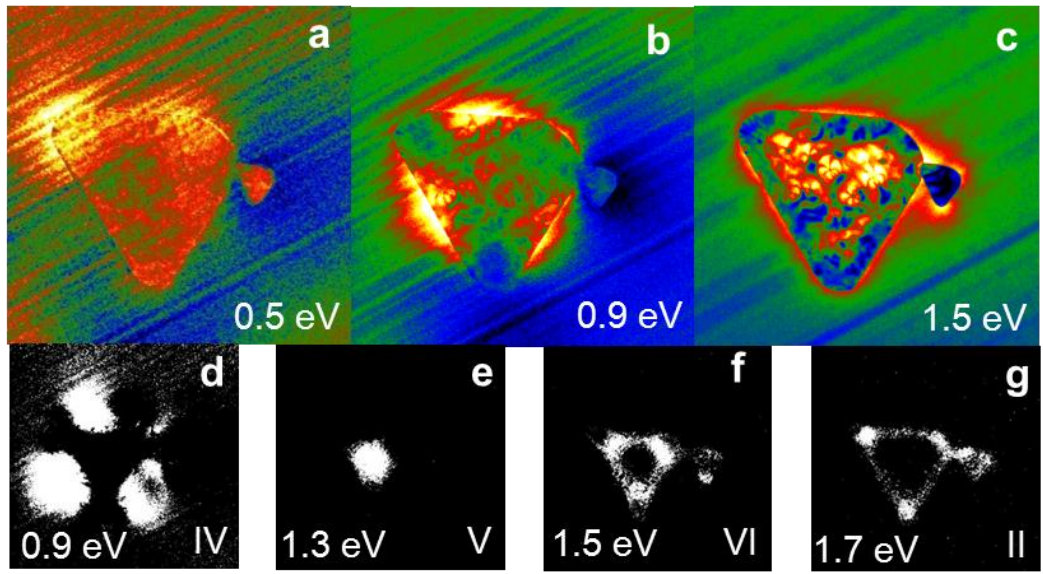

Figure 8: (a–c) EFTEM images of a small triangular Au nanoprism positioned close to the larger Au nanoprism, near to the upper right corner. (d–g) peak maps at different energy losses, extracted from the EFTEM series. Roman numbers denote the modes as specified in Figure 7.

## CONCLUSIONS

We presented here a systematic study of the symmetry-breaking effect on the eigenmodes of a gold triangular nanoprism. We showed that breaking the symmetry of the structure results in splitting the degeneracy of the eigenmodes and even exchanging the sequence of eigenenergies regarding the incident polarization. Moreover, an induced dipolar mode of the nanoprism is observed at a much higher frequency than the corresponding mode of the simple nanoprism. Using the decomposition techniques, the different modes of the nanoprisms were investigated according to their symmetry. It has been shown that due to the interference of a broad monopolar mode with a narrow multipolar mode, a Fano-type resonance is observable at the near-field spectrum of a single nanoprism. In the broken symmetry case, this resonance rapidly deteriorates.

## ACKNOWLEDGMENT

We would like to thank Dr. Isabel Pastoriza-Santos for her advice regarding the choice of the synthesis method. N. Talebi would like to thank the Alexander-von-Humboldt Foundation for the financial support. L.M.L.-M. acknowledges support from Xunta de Galicia (grant #09TMT011314PR).

# REFERENCES

[1] Nelayah, J.; Kociak, M.; Stéphan, O.; Garcia de Abajo, F. J.; Tencé, M.; Henrard, L.; Taverna, D.; Pastoriza-Santos, I.; Liz-Marzán, L. M.; Colliex, C. *Nature Physics* **2008**, *3*, 348–353.

[2] Shuford, K. L.; Ratner, M. A.; Schatz, G. C. *J. Chem. Phys.* **2005**, *123*, 114713-1–9.

[3] Nelayah, J; Kociak, M.; Stéphan, O.; Geuquet, N.; Henrard, L.; Garcia de Abajo, F. J.; Pastoriza-Santos, I.; Liz-Marzán, L. M.; Colliex, C. *Nano Lett.* **2010**, *10*, 902–907.

[4] Gu, L.; Sigle, W.; Koch, C. T.; Ögüt, B.; van Aken, P. A.; Talebi, N.; Vogelgesang, R.; Mu, J.; Wen, X.; Mao, J. *Phys. Rev. B* **2011**, *83*, 195433-1–7.

[5] Wang, H.; Wu, Y.; Lassiter, B.; Nehl, C. L.; Hafner, J. H.; Nordlander, P.; Halas, N. J. *PNAS* **2006**, *103*, 10856–10860.

[6] Rang, M.; Jones, A. C.; Zhou, F.; Li, Z.-Y.; Wiley, B. J.; Xia, Y.; Raschke, M. B. *Nano Lett*. **2008**, *8*, 3357–3363.

[7] Lassiter, J. B.; Sobhani, H.; Fan, J. A.; Kundun, J.; Capasso, F.; Nordlander, P.; Halas, N. J. Fano Resonances in Plasmonic Nanoclusters: Geometrical and Chemical Tunability. *Nano Lett.* **2010**, 10, 3184-3189.

[8] Hentschel, M.; Saliba, M.; Vogelgesang, R.; Giessen, H.; Alivisatos, A. P.; Liu, N. Transition from isolated to collective modes in plasmonic oligomers. *Nano Lett.* **2010**, 10, 2721-2726.

[9] Hentschel, M.; Dregely, D.; Vogelgesang, R.; Giessen, H.; Liu, N.; Plasmonic Oligomers: The Role of Individual Particles in Collective Behavior. *ACS Nano* **2011**, 5, 2042-2050

[10] Halas, N. J.; Lal, S.; Chang, W.; Link, S.; Nordlander, P. *Chem. Rev.* **2011**, *111*, 3913-3961.

[11] Millstone, J. E.; Wei, W.; Jones, M. R.; Yoo, H.; Mirkin, C. A. *Nano Lett*. **2008**, *8*, 2526–2529.

[12] Koch, C. T.; Sigle, W.; Höschen, R.; Rühle, M.; Essers, E.; Benner, G.; Matijevic, M. *Microsc. Microanal.* **2006**, *12*, 506–514.

[13] Uhlemann, S.; Rose, H. *Ultramicroscopy* **1996**, *63*, 161–167.

[14] Egerton, R. F.; Electron Energy-Loss Spectroscopy in the Electron Microscope, 2d Edition, Plenum Press, New York, **1996**, 347–352.

[15] Etchegoin, P. G.; Le Ru, E. C.; Meyer, M. *J. Chem. Phys.* **2006**, *125*, 1564705.

[16] Harris, D. C. and Bertolucci, M. D., "Symmetry and Spectroscopy: Introduction to Vibrational and Electronic Spectroscopy", Oxford University Press, (1980).

[17] Bishop, D. M., "Group Theory and Chemistry", Dover Publications, (1993).

[18] Lukyanchuk, B; Zheluev, N. I.; Maier, S. A.; Halas, N. J.; Nordlander, P; Giessen, H., Chong, C. T. *Nautre Materials*, **2010**, *9*, 707-715.